\pdfoutput=1

\documentclass[11pt]{article}
\usepackage{caption}
\usepackage{enumitem}
\usepackage{soul,color}
\definecolor{burgundy}{rgb}{0.5, 0.0, 0.13}
\definecolor{dgreen}{rgb}{0.0, 0.61, 0.0}
\definecolor{aqua}{cmyk}{0.91, 0, 0.09, 0.36}
\definecolor{aqua}{cmyk}{0.91, 0, 0.09, 0.36}
\definecolor{dblue}{cmyk}{0.999, 0.33, 0.1, 0.15}

\usepackage{acl}
\usepackage{amstext} 
\usepackage{array}   
\newcolumntype{L}{>{$}l<{$}} 
\newcolumntype{C}{>{$}l<{$}}
\usepackage{times}
\usepackage{latexsym}
\usepackage{booktabs}
\usepackage{multirow}
\usepackage{graphicx}

\usepackage[T1]{fontenc}

\usepackage[utf8]{inputenc}

\usepackage{microtype}

\usepackage{inconsolata}

\usepackage{booktabs}
\usepackage{multirow}
\usepackage{array}
\usepackage{varwidth}
\newcolumntype{M}{>{\begin{varwidth}{16cm}}l<{\end{varwidth}}} 
\newcolumntype{E}{>{\begin{varwidth}{8cm}}l<{\end{varwidth}}} 
\newcolumntype{N}{>{\begin{varwidth}{5cm}}l<{\end{varwidth}}} 
\newcolumntype{P}[1]{>{\centering\arraybackslash}p{#1}}
\newcommand{\generation}[1]{#1{\color[HTML]{CB4335}$\dagger$}}
\title{Social Intelligence Data Infrastructure: \\Structuring the Present and Navigating the Future}

\newcommand{\treelogo}{\raisebox{5pt}{\includegraphics[scale=0.050]{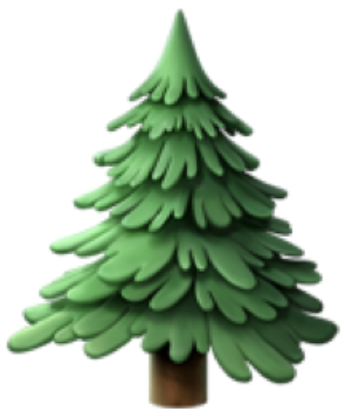}}}
\newcommand{\lionlogo}{\raisebox{5pt}{\includegraphics[scale=0.040]{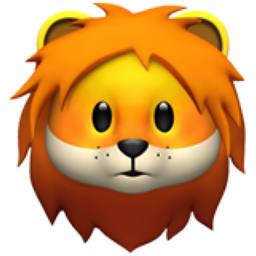}}}
\newcommand{\starlogo}{\raisebox{5pt}{\includegraphics[scale=0.040]{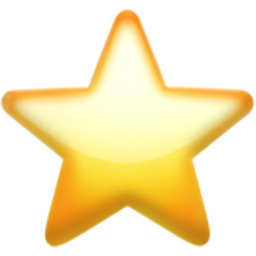}}}

\newcommand{\nus}{\lionlogo}
\newcommand{\astar}{\starlogo}
\newcommand{\stanford}{\treelogo}

\author{Minzhi Li \nus \astar \hspace{1.5em}
        Weiyan Shi \stanford \hspace{1.5em}
        Caleb Ziems \stanford \hspace{1.5em}
        \bf Diyi Yang \stanford \hspace{1.5em}\\
        \nus National University of Singapore \hspace{1.5em}\\
        \astar Institute for Infocomm Research (I$^2$R), A*STAR \\
        \stanford Stanford University\\
        \texttt{\href{mailto://li.minzhi@u.nus.edu}{li.minzhi@u.nus.edu}} \hspace{1.5em}
        \texttt{\href{mailto://weiyans@stanford.edu}{weiyans@stanford.edu}}  \\ 
        \texttt{\href{mailto://cziems@stanford.edu}{cziems@stanford.edu}} \hspace{1.5em}
        \texttt{\href{mailto://diyiy@cs.stanford.edu}{diyiy@cs.stanford.edu}} 
}

\begin{document}
\maketitle
\begin{abstract}
As Natural Language Processing (NLP) systems become increasingly integrated into human social life, these technologies will need to increasingly rely on social intelligence. Although there are many valuable datasets that benchmark isolated dimensions of social intelligence, there does not yet exist any body of work to join these threads into a cohesive subfield in which researchers can quickly identify research gaps and future directions. Towards this goal, we build a \textit{Social AI Data Infrastructure}, which consists of a comprehensive social AI taxonomy and a data library of 480 NLP datasets. Our infrastructure allows us to analyze existing dataset efforts, and also evaluate language models' performance in different social intelligence aspects. Our analyses demonstrate its utility in enabling a thorough understanding of current data landscape and providing a holistic perspective on potential directions for future dataset development. We show there is a need for multifaceted datasets, increased diversity in language and culture, more long-tailed social situations, and more interactive data in future social intelligence data efforts.
\end{abstract}

\section{Introduction}

\begin{quote}
    {
    ``\textit{Data is a precious thing and will last longer than the systems themselves.}'' \\\phantom{abc}--- \textbf{Tim Berners-Lee} 
    
    }
\end{quote}

As early as the 1920s, psychologists like \citet{thorndike1921intelligence} and \citet{hunt1928measurement} considered social intelligence to be a distinct branch of intelligence that underlies all successful human interpersonal relationships.
Many researchers now argue that social intelligence is a prerequisite of human-like Artificial Intelligence \citep{kihlstrom2000social, erickson2009social, del2019you, radfar2020characterizing, hovy2021importance, williams2022supporting}. However, existing work still lacks a precise yet holistic definition for social intelligence in AI systems
\citep{silvera2001tromso}. 

\begin{figure}[t!]
    \centering \includegraphics[width=\linewidth]{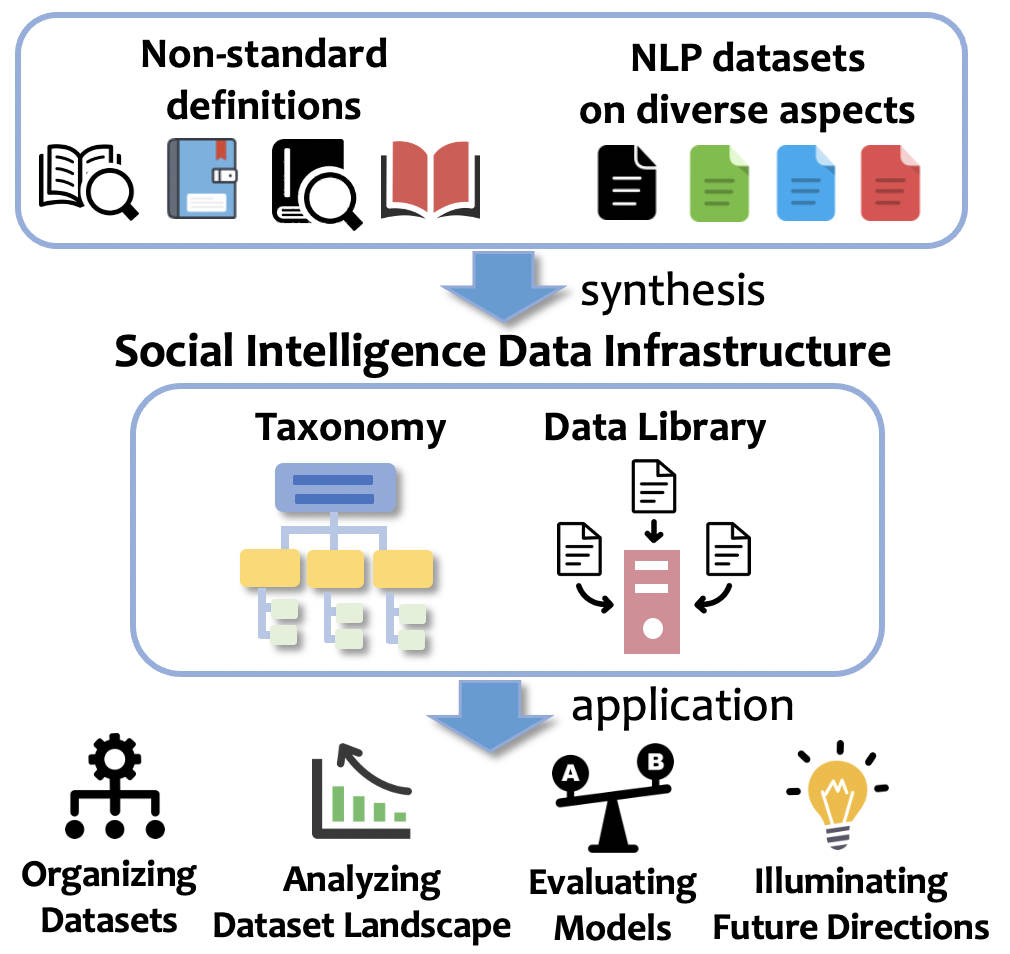}
    \caption{Our \textit{Social Intelligence Data Infrastructure} gives a comprehensive overview and synthesis of social intelligence in NLP, with a theoretically grounded taxonomy and an NLP data library. Researchers can use our infrastructure to build and organize tasks, evaluate language models and derive future insights.}\vspace{-0.8em} 
    \label{fig:crown_jewel}
\end{figure}

Prior studies define social intelligence by its cognitive aspect, or the ability to \textit{understand} others \citep{barnes1989social},  while others emphasize the behavioral component by defining it as the ability to \textit{interact} with other people \citep{hunt1928measurement, ford1983further}. Both dimensions are relevant but incomplete, as social intelligence is multi-faceted \citep{marlowe1986social}. Besides the narrow definitions, related empirical efforts, such as dataset collection and model development, also have isolated focuses and only address a single aspect of social intelligence \citep{fan2022artificial}. Therefore, there exists a pressing need for a holistic and synthesized definition for social intelligence to create an organized space for existing datasets. Without such organization, it is difficult to identify overarching research questions and emerging trends for future exploration. Besides, although various social intelligence datasets have been proposed, the lack of data organization creates barriers for researchers to gain insights from previous work. 

In light of this, we establish \textit{Social Intelligence Data Infrastructure}, which consists of a comprehensive taxonomy for social intelligence and an organized data library of 480 NLP datasets (see Figure \ref{fig:crown_jewel}) to structure current data efforts and navigate future directions. The taxonomy (\S \ref{sec:taxonomy}) formally defines various aspects of social intelligence, to introduce standardization and comprehensiveness to the definition of social intelligence in AI systems. The data library (\S \ref{sec:library}) maps crawled datasets to different categories in our taxonomy, to provide structures for existing datasets. The taxonomy and data library can collectively aid researchers to identify existing dataset gaps and guide future dataset development for social intelligence.

Moreover, we demonstrate how the proposed \textit{Social Intelligence Data Infrastructure} can be applied to gain insights into future dataset development with the following contributions:
\begin{itemize}\itemsep0em
\item We perform distributional and temporal analysis (\S \ref{sec:analysis}) to highlight overlooked categories and uncover emerging trends. 
\item We evaluate the zero-shot performance of Large Language Models (LLMs) (\S \ref{sec:model}) on various social intelligence aspects defined in our taxonomy, to shed light on current models' capabilities and limitations. 
\item Finally, guided by the analysis and evaluation results,  we discuss unfilled gaps and future directions for NLP dataset efforts on social intelligence (\S \ref{sec:insights}). We identify a need for multifaceted datasets, better diversity in language and culture, more long-tailed social situations, and more interactive data. 

\end{itemize}

\begin{figure*}[t]
  \includegraphics[width=\textwidth]{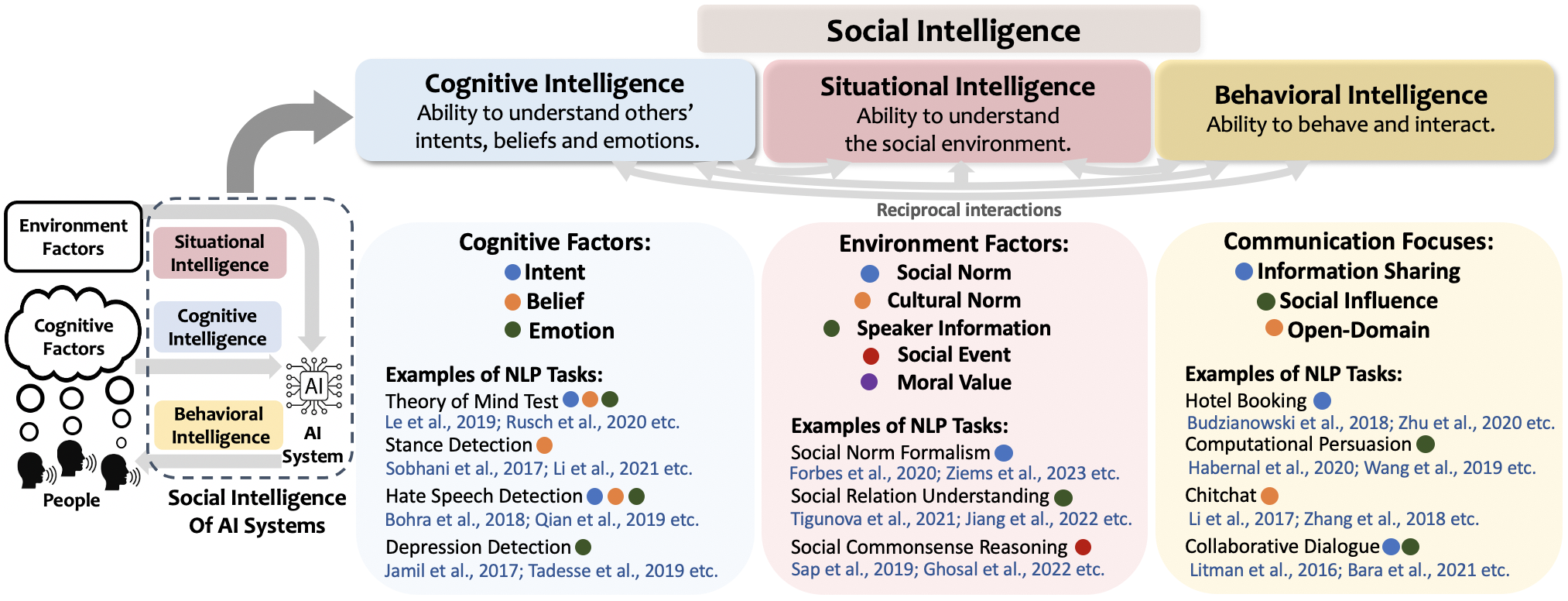}
  \caption{Social AI taxonomy with three pillars: cognitive, situational and behavioral intelligence. We illustrate their respective roles in social interactions (left), and visualize their definitions and example NLP tasks (right).}
  \label{fig:taxonomy}
\end{figure*}

\section{Social AI Taxonomy}
\label{sec:taxonomy}

To introduce a standardized and comprehensive definition of social intelligence, we propose \textit{Social AI Taxonomy}, to capture diverse 
dimensions identified in previous work. As shown in Figure \ref{fig:taxonomy}, different from previous categorization which is thematic with a focus on social understanding \citep{choi2023llms}, our taxonomy considers the social interaction component and is hierarchical with three distinct types of social intelligence based on past literature: (1) cognitive intelligence, (2) situational intelligence, and (3) behavioral intelligence.   

As identified in social cognitive theory \citep{bandura2009social}, these three intelligence types mutually influence each other to shape human behaviour. Prior work \citep{barnes1989social,kosmitzki1993implicit} shows that these three intelligence types can comprehensively cover different factors and dimensions under social intelligence. Now we detail each intelligence type below.

\subsection{Cognitive Intelligence}
\label{subsec:cognitive_intelligence}
\noindent\textbf{Definition.} Cognitive intelligence refers to the use of verbal and nonverbal cues \citep{hunt1928measurement} to understand others' mental states \citep{barnes1989social}. These include cold cognition about intents and beliefs, as well as hot cognition about emotions \citep{roiser2013hot}, so our taxonomy decomposes cognitive intelligence into knowledge about \textit{\textbf{intents}}, \textit{\textbf{beliefs}}, and \textit{\textbf{emotions}}.

\textbf{Significance.} Cognitive intelligence includes the prerequisites for effective communication \citep{apperly2010mindreaders} and many concrete NLP tasks \citep{langley2022theory}. Task-oriented dialogue requires intent recognition \citep{jayarao2018intent, wu2023joint}, and mental health support demands an understanding of emotions \citep{peng2021research, singh2023emotion}. 
Broadly, Theory of Mind is a fundamental module in both both human \citep{premack1978does} and artificial social intelligence \citep{rusch2020theory} that underlies downstream skills like stance awareness (see the left pillar of Figure \ref{fig:taxonomy}).

\subsection{Situational Intelligence}
\noindent\textbf{Definition.} Situational intelligence refers to an awareness of the social context \citep{derks2007emoticons} and how this context informs the other pillars of cognition and behavior (\textit{reciprocal interactions} in Figure \ref{fig:taxonomy}). The literature tells us that social context includes: \textit{\textbf{the social event}} itself \citep{sap2019socialiqa}, as well as \textit{\textbf{social and moral norms}} \citep{ziems-etal-2023-normbank}, \textit{\textbf{culture}} and \textit{\textbf{speaker information}} \citep{tigunova2021pride}, all included in the \textit{Taxonomy}.

\textbf{Significance.} Situational intelligence makes use of social context as the glue to bind the cognitive intelligence of mental states (\S\ref{subsec:cognitive_intelligence}) with a set of appropriate behaviors (\S\ref{subsec:behavioral_intelligence}), and serves as the foundation for decision making \citep{endsley1990situation}. Given its centrality and its clear manifestations in decision making, situational intelligence has been a standard locus for social intelligence tests \citep{lievens2017practical, hunt1928measurement}.
The social context spans not only interpersonal factors like tie strength \citep{sap2019risk, cristani2011towards}, but also cultural differences like those between high and low-context cultures \citep{hall1987hidden, devito2016interpersonal}, and navigating these differences is essential for cross-domain and cross-cultural communication \citep{wawra2013social}.
Studies have shown that incorporating these factors can lead to significant performance improvements in NLP systems \citep{rahimi-etal-2018-semi, wu2021personalized}. \vspace{-0.05in}
\subsection{Behavioral Intelligence}
\label{subsec:behavioral_intelligence}
\noindent\textbf{Definition.} Behavioral intelligence refers to skills of successfully communicating and acting in a manner to attain social goals \citep{ford1983further} through (1) \textbf{information sharing} \citep{zhang2020recent}, (2) \textbf{social influence} \citep{turner1991social, cialdini2004social, chawla2023social,weinstein1969development}, or (3) maintaining interpersonal relationships \citep{vernon1933some,moss1927you} via \textbf{open-domain} conversations \citep{huang2020challenges}. 
Our \textit{Taxonomy} is organized around these three foci.

\textbf{Significance.} Behavioral intelligence has direct ramifications for human-human and human-AI interactions (see the right pillar of Figure~\ref{fig:taxonomy}). Task-oriented dialogue systems \citep{zhang2020recent} and collaborative AI partners \citep{bara2021mindcraft} depend on successful information sharing, while other applications require engaging and personalized open-domain chit-chat \citep{zhang-etal-2018-personalizing}. The capacity for social influence becomes relevant in human-AI teams \citep{bansal2021does}, where such skills can make use of advances in explainable AI systems \citep{angelov2021explainable}. Across these diverse applications, systems need to be equipped with social-behavioral skills like empathy \citep{rashkin2018towards}, persuasion \citep{hunter2019towards}, and transparency to build trust \citep{liao2022designing}.

\vspace{-0.1in}
\subsection{Challenges in Measuring Intelligence}
The three pillars of social intelligence are \textit{not} mutually exclusive, nor are they readily isolated in social life, since they coordinate through reciprocal interactions (Figure \ref{fig:taxonomy}). A situationally intelligent agent can better express cognitive intelligence, using cues from the social context to infer the mental states of others. The converse is also true that, by considering others' mental states, one can understand her role in shaping the social situation. Both intelligence can facilitate effective social actions.

Because of its dynamic nature, social intelligence may be outside the scope of what AI engineers can benchmark with any single static dataset. This is especially true as \textit{situational} and \textit{behavioral} pillars themselves are not static, but refer rather to an agent's ability to adapt into a social equilibrium. In our analysis and discussion, we will consider the degree to which benchmarks can reflect this dynamic nature, and whether existing datasets measure more than one type of social intelligence. For example, the \textsc{CobraCorpus} \citep{zhou2023cobra} requires cognitive and situational intelligence to reason about offensive intents in different social contexts. Finally, we make suggestions for the design of data resources and the future of social AI.

\section{Current Social NLP Data Landscape}
With what granularity can existing data resources help researchers train and evaluate the core pillars of social intelligence in AI systems? How holistic is the landscape, and how sufficiently integrated are these pillars in the literature? To answer these questions, we leverage the \textit{Social AI Taxonomy} to categorize existing NLP publications into a library of relevant datasets. 
\vspace{-0.7em}
\subsection{Data Library Construction}
\label{sec:library}
We use ACL Anthology data\footnote{\url{https://aclanthology.org/}} crawled by \citeauthor{rohatgi-etal-2023-acl} and \citet{held2023material}, 
and set our time scope from year 2001 January to 2023 October as there are few datasets before 2001. We automatically collect social intelligence dataset papers by filtering titles and abstracts with keywords related to both (a) social intelligence and (b) dataset development (see Appendix \ref{sec:keyword}). The smaller size of this filtered pool allows us to manually curate papers, removing any surveys or irrelevant works on model development, annotation schemes, or annotation tools, which results in a curated set of 480 papers. For these papers, we scraped useful metadata, like title, url and publication year. 
We discuss the technically infeasibility of an exhaustive library in \S \ref{sec:limitations}.
\vspace{-0.7em}
\subsection{Metadata Annotation}
We map the papers in our data library to the \textit{Social AI Taxonomy}. Two authors reviewed the content in the paper and annotated them with type of intelligence and \S\ref{sec:taxonomy} subcategory (Cohen $\kappa=$ 0.86 \textit{cognitive}; 0.80 \textit{situational}; 0.87 \textit{behavioral}). The annotation is based on the main focus of the dataset. For example, if a work collects interactive dialogues solely for intent recognition purpose, we will classify it as cognitive intelligence instead of behavioral intelligence. Constructing the data library \footnote{\href{https://docs.google.com/spreadsheets/d/1jSTmPaqaEVXxoLmt_DCk933PsthMucbIFT7KmZt2Q3A/edit?usp=sharing}{Data library}}  illustrates how our theoretical taxonomy can be practically useful to organize datasets focusing on different aspects of social intelligence. On top of that, we also annotate other important attributes for each dataset by reviewing the paper contents:  we annotate \textit{\textbf{NLP Task}}, \textit{\textbf{Data Source}} (where the data was collected from), \textit{\textbf{Annotation Strategy}} (how the labels were obtained), \textit{\textbf{Generation Method}} (if the text comes from human, AI or both), \textit{\textbf{Data Format}} (e.g. tweet, news article, dialogue etc.), \textit{\textbf{Language}}, \textit{\textbf{Modality}}, and \textit{\textbf{Public Availability}} of the data.

\subsection{Social NLP Data Landscape Highlights}
\label{sec:analysis}
By visualizing the distribution and temporal trend of the datasets in the data library, we obtain insights about the past and current NLP paradigm for dataset development on social intelligence. We discuss key results in this section and put more detailed analyses in Appendix \ref{sec:appanalysis}.

\begin{figure}[t]
    \centering \includegraphics[width=\linewidth]{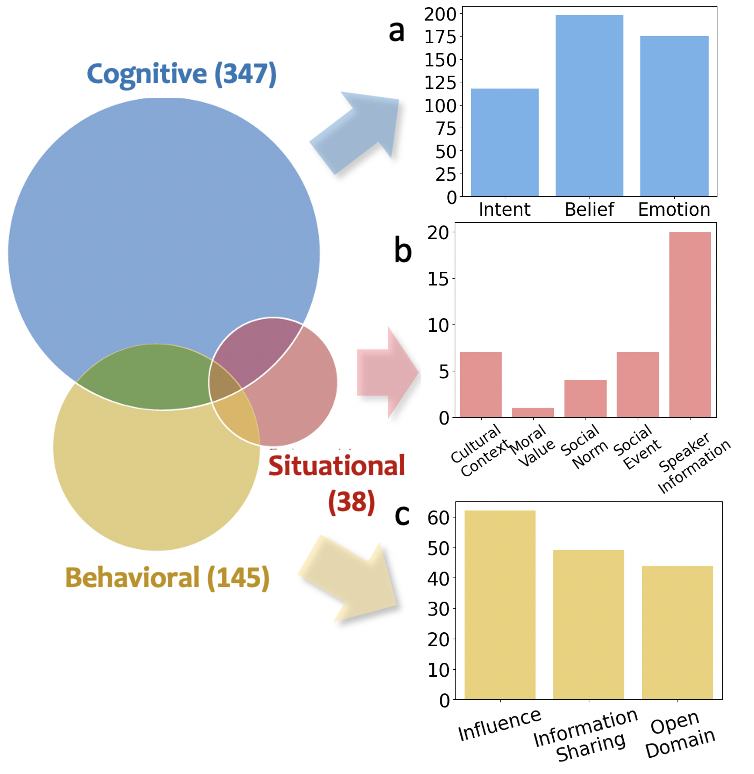}
    \caption{Distribution of three intelligence types (left) and frequency of different subcategories within cognitive, situational and behavioral intelligence (right).} \vspace{-1em}
    \label{fig:in_type}
\end{figure}

\begin{figure*}[h]
  \includegraphics[width=\textwidth]{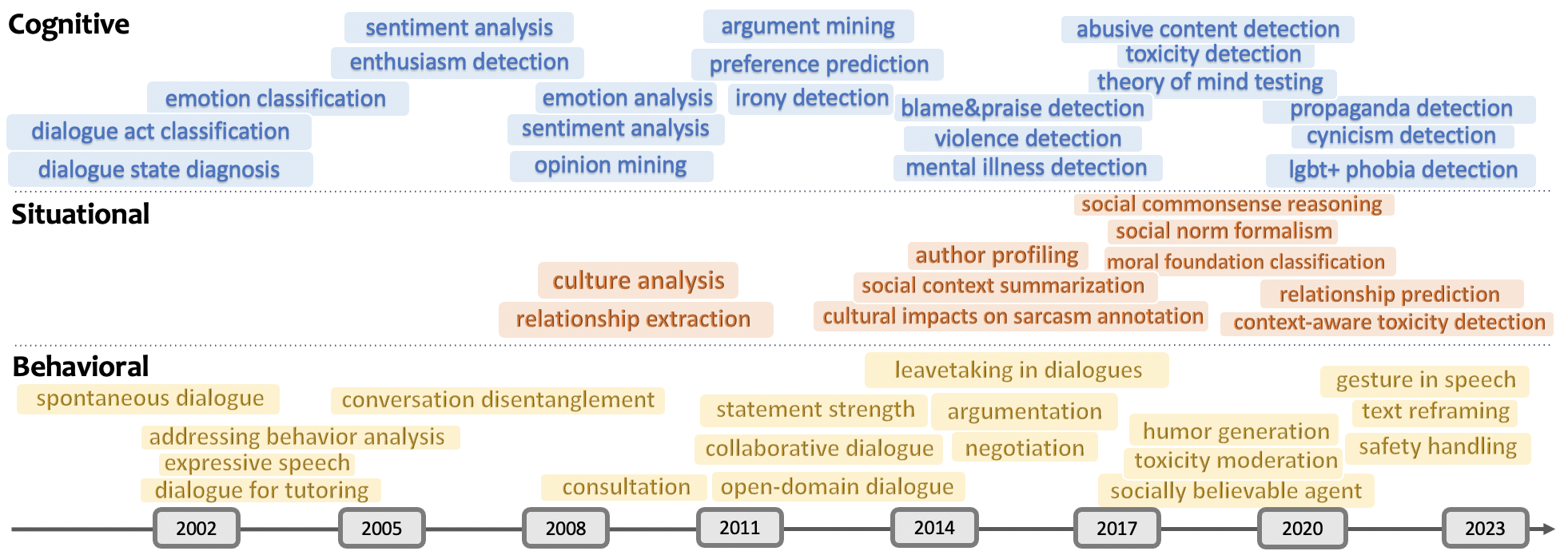}
  \caption{NLP tasks related to social intelligence over time. We show newly emerged topics based on the NLP Task field in our constructed data library for every three years. This is a non-exhaustive visualization (if number of distinct new topics for the period is more than three, we cap at three).} 
  \label{fig:temporal}
\end{figure*}

\paragraph{Topic Distribution} Figure \ref{fig:in_type} shows most of the social NLP datasets focus on the cognitive aspect of social intelligence (64.2\%), followed by behavioral aspect (22.7\%), and least of all, its situational aspect (3.8\%). Only a \textit{\textbf{small set}} of datasets (9.4\%) span \textit{\textbf{multiple intelligence types}}. We also visualize a detailed breakdown of different factors within each type. For cognitive and behavioral 
intelligence, papers are balanced across the respective subcategories. For situational intelligence, most datasets measure knowledge of speakers involved in the dialogue such as their demographics and social relations, and there are \textit{\textbf{very few datasets on moral values and social norms}}.

\paragraph{Temporal Topic Shift} From Figure \ref{fig:temporal}, we can better understand the temporal variation and shift of focus over time for each type of intelligence. 
We can see the \textit{\textbf{onset of study on situational intelligence is later}} (2008) than the other two types (2001). For all three intelligence, the task of focus has become more \textit{\textbf{specific and nuanced}} over the years. For example, early work on cognitive intelligence focused on general dialogue act classification  but recent studies are about more nuanced and challenging intents beyond literal meaning like sarcasm and irony understanding \citep{alnajjar2021qu, frenda2023epic}. Literature on behavioral intelligence began with tasks to identify effective or powerful written communication \citep{tan2014corpus}, and moved to more specific tasks like high-quality persuasive arguments for particular forms of negotiation \citep{ng2020creating, chawla2021casino}.

\paragraph{Interactive vs Static Data }
\begin{figure}[h!]
    \centering \includegraphics[width=\linewidth]{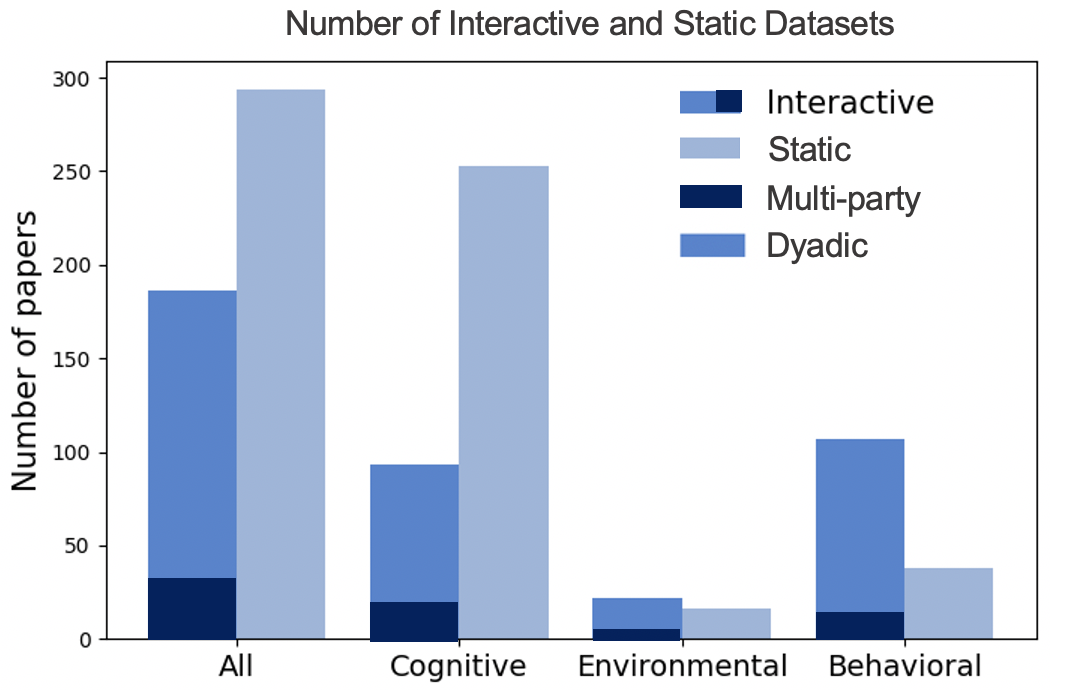}
    \caption{Number of papers with interactive or static data. We also visualize a breakdown of interactive data into dyadic and multi-party interactions.}
    \label{fig:format}
\end{figure}
We classify the data formats of surveyed datasets into two broad categories -- interactive and static. Interactive data are those with information exchange like social media threads and daily conversations. Static data include (1) self-contained and topically focused texts written for a general audience (e.g. news and books) and (2) those that are part of information exchange but  have no prior or subsequent context (e.g. a single post on Twitter or an utterance in a conversation). The difference in data format can affect the ability for language model to acquire social intelligence \citep{sap2022neural}. Figure \ref{fig:format} shows that \textbf{\textit{for cognitive aspects}} in particular, but not for other pillars, there are significantly \textit{\textbf{more static datasets}} than interactive ones, which quantitatively confirms the trend observed by \citet{sap2022neural}. Moreover, within the interactive datasets, the \textit{\textbf{proportion of multi-party modeling is small}} (18.4\%).

\paragraph{Use of AI} There are \textit{\textbf{increasing number of works adopting AI}} for generating and annotating datasets related to social intelligence (before 2015: 3; after 2015: 32). We find that \textit{\textbf{degree of adoption of AI for generation is higher}} than annotation. In recent work, researchers outsource generation completely by generating contents purely using AI \citep{zhou2023cobra} or simulating conversations between AI and AI \citep{lee2022personachatgen}. On the other hand, use of AI in annotating social intelligence data still remains in a hybrid stage \citep{jo2020machine} and AI usually plays a part in annotating simpler high-level components like themes \citep{maes2023studying}. 
\section{Model Performance}

Now we use our \textit{Social AI Data Infrastructure} to evaluate current LLMs' performance on social intelligence and gain insights on models' strengths and limitations, shedding light on aspects which future social intelligence datasets need to address. 
\label{sec:model}
\begin{table*}[t]
\centering
\def\arraystretch{1.8}
\resizebox{\textwidth}{!}{  
\begin{tabular}{llllCCCCC}
\hline
\multirow{2}{*}{\textbf{Intelligence}} & \multirow{2}{*}{\textbf{Category}}                                       & \multirow{2}{*}{\textbf{Dataset \small(year)}} & \multirow{2}{*}{\textbf{Task}}                    & \multicolumn{2}{c}{\textbf{Human}} & \multicolumn{3}{c}{\textbf{LLM}}       \\ \cline{5-9} 
                                       &                                                                             &                                          &                                                   & Average       & Best               & Claude & GPT-4         & Llama2    \\ \hline
\multirow{6}{*}{Cognitive}             & \multirow{2}{*}{Intent}                                                     & \href{https://huggingface.co/datasets/snips_built_in_intents}{SNIPS} \small(2018)                             & query intent classification                       & 82.5          & \textbf{97.5}      & 78.8   & 95.0^*          & 76.7          \\ \cline{3-9} 
                                       &                                                                             & \href{https://github.com/dmbavkar/iSarcasm/blob/master/isarcasm_test.csv}{iSarcasm} \small(2018)                           & intended sarcasm detection                        & 64.8          & \textbf{90.9}      & 56.4   & 67.3^*          & 55.5          \\ \cline{2-9} 
                                       & \multirow{2}{*}{Belief}                                                     & \href{https://github.com/cardiffnlp/tweeteval/tree/main/datasets/stance/abortion}{SemEvalT6} \small(2016)                         & stance detection on abortion                      & 54.4          & \textbf{84.2}      & 59.2^*   & 76.7^*          & 45.8          \\ \cline{3-9} 
                                       &                                                                             & \href{https://github.com/cambridge-wtwt/acl2020-wtwt-tweets}{WTWT} \small(2020)                              & stance detection on merger and acquisition        & 61.4          & \textbf{81.7}      & 47.5   & 55.8          & 75.0^*          \\ \cline{2-9} 
                                       & \multirow{2}{*}{Emotion}                                                    & \href{https://huggingface.co/datasets/sem_eval_2018_task_1/viewer/subtask5.english/test}{SemEvalT1} \small(2018)                         & emotion classification (11 classes)               & 78.9          & \textbf{83.2}      & 78.5   & 80.9^*         & 78.6          \\ \cline{3-9} 
                                       &                                                                             & \href{https://huggingface.co/datasets/go_emotions}{GoEmotions} \small(2020)                             & emotion classification (28 classes)               & 90.9          & \textbf{93.1}      & 92.0^*   & 90.2          & 92.6^*          \\ \hline
\multirow{4}{*}{Situational}         & \multirow{2}{*}{\begin{tabular}[c]{@{}c@{}}Social\\ Situation\end{tabular}} & \href{https://huggingface.co/datasets/social_i_qa}{SocialIQa} \small(2019)                         & commonsense reasoning with description            & 55.8          & \textbf{80.0}      & 70.8^*   & 70.8^*          & 63.3^*          \\ \cline{3-9} 
                                       &                                                                             & \href{https://declare-lab.github.io/CICERO/}{CICERO} \small(2022)                            & commonsense reasoning with dialogues              & 77.2          & 85.3               & 79.3^*   & \textbf{86.7}^* & 76.2          \\ \cline{2-9} 
                                       & \multirow{2}{*}{\begin{tabular}[c]{@{}c@{}}Social\\ Norm\end{tabular}}      & \href{https://drive.google.com/drive/folders/1XRhrzgG_R0zypPgPlCxK0nlqKbfaI9xe}{NormBank} \small(2023)                          & situational judgment                              & 52.2          & \textbf{71.7}      & 52.5^*   & 60.8^*          & 40.0          \\ \cline{3-9} 
                                       &                                                                             & \href{https://huggingface.co/datasets/feradauto/MoralExceptQA}{MoralExceptQA} \small(2022)                     & situational exception judgment                    & 44.8          & \textbf{93.2}      & 47.3^*   & 50.0^*          & 33.1          \\ \hline
\multirow{6}{*}{Behavioral}           & \multirow{2}{*}{ChitChat}                                                   & \href{https://huggingface.co/datasets/daily_dialog}{DailyDialogue} \small(2017)                     & daily conversations\generation{}                               & 50.0          & $-$                  & 74.7^*   & 84.4^*          & \textbf{88.6}^* \\ \cline{3-9} 
                                       &                                                                             & \href{https://huggingface.co/datasets/AlekseyKorshuk/persona-chat}{PersonaChat} \small(2018)                       & chats conditioned on personas \generation{}                    & 50.0          & $-$                  & 64.0^*   & \textbf{78.5}^* & 64.0^*          \\ \cline{2-9} 
                                       & \multirow{2}{*}{Persuasion}                                                 & \href{https://github.com/UKPLab/emnlp2016-empirical-convincingness/blob/5396e5ae06dd65c064fc0864f106f095f47acfe7/data/CSV-format/ban-plastic-water-bottles_yes-emergencies-only.xml.csv}{Convincing Arguments} \small(2016)              & argument generation \generation{}                              & 50.0          & $-$                  & 93.6^*   & \textbf{97.8}^* & 93.1^*          \\ \cline{3-9} 
                                       &                                                                             & \href{https://gitlab.com/ucdavisnlp/persuasionforgood/-/blob/master/data/AnnotatedData/300_dialog.xlsx?ref_type=heads}{PersuasionforGood} \small(2019)                 & persuasive dialogue response generation \generation{}          & 50.0          & $-$                  & 74.3^*   & \textbf{96.3}^* & 93.9^*           \\ \cline{2-9} 
                                       & \multirow{2}{*}{Therapy}                                                    & \href{https://github.com/SALT-NLP/positive-frames/blob/main/data/wholetest.csv}{Positive Reframing} \small(2022)                & reframing text in a positive way \generation{}                  & 50.0          & $-$                  & 83.5^*   & \textbf{96.2}^* & 92.0^*          \\ \cline{3-9} 
                                       &                                                                             & \href{https://huggingface.co/datasets/nbertagnolli/counsel-chat}{Counsel-Chat} \small(2020)                      & provide counseling to problems  \generation{}                  & 50.0          & $-$                  & 62.6^*   & \textbf{93.5}^* & 84.6^*          \\ \hline
\multirow{2}{*}{Multiple}              & \small Intent+Social Situation                                                     & \href{https://huggingface.co/datasets/cmu-lti/cobracorpus}{\textsc{CobraCorpus}} \small(2023)                                    & contextual offensive statement detection          & 73.6          & \textbf{95.0}      & 70.0   & 89.2^*          & 50.0          \\ \cline{2-9} 
                                       & \small Intent+ Cultural Norm                                                        & \href{https://github.com/SALT-NLP/CulturallyAwareNLI/blob/main/data/data.tsv}{CulturalNLI} \small(2023)                              & culturally aware natural language inference & 52.1          & \textbf{72.3}      & 33.3   & 65.0^*          & 41.7          \\ \hline
\end{tabular}}
\caption{Human and LLMs' performance on classification (F1 scores) and generation tasks (\% preferred over average human). Within each category, we select one simpler (top) and one more nuanced (bottom) dataset for comparison. LLM performance that exceeds average human performance is marked with $^*$ and best performance is \textbf{bolded}. \generation{} Generation tasks have average human performance of 50\% by definition; best performance is not defined.} 
\label{tab:result}
\end{table*}

\subsection{Experimental Setup}

\noindent\textbf{Dataset selection}.
For each of our taxonomic categories, we select two representative datasets, one simple and the other more challenging. Challenges arise from factors like nuancedness (e.g., sarcastic intents), task granularity (e.g., emotion detection with more fine-grained classes), data scarcity (e.g., stance detection in the economic domain), long-tail data distributions (e.g., moral exception), and challenging data formats (e.g., persuasion in a dialogue setting). We select open-sourced and widely-used  datasets with high citations for each category. Testing the model on the entire dataset can be computationally expensive and time-consuming so we adopt class-stratified sampling of 100 to 150 instances from the original test set (if available). We evaluate LLMs' zero-shot performance on sampled instances, and follow the recommended practice of prompting as described by \citet{ziems2023can}.

\noindent\textbf{Metrics.} 
For classification tasks, we choose F1 as the metrics. For generation tasks, we present both the original human response and the LLM response to human annotators, and calculate the preference percentage.  As such, we can better understand current social capability of language models, in both absolute terms against the ground truth and relative terms compared to human.

\noindent\textbf{Human performance.} For classification, we report the best and average F1 scores for responses from three MTurk workers per test instance. For generation tasks in which we compute preferences, we define the average human performance as 50\%, without defining the best human performance.

\subsection{Result Analysis}
\noindent\textbf{Main Results.} From Table \ref{tab:result}, we can see that \textit{\textbf{LLM performs better on simpler datasets than more nuanced ones}}. For example, compared to straightforward query intent recognition (95.0 F1), the best performing LLM (GPT-4) struggles more with identifying the intended sarcasm (67.3 F1) when people convey an opposite meaning from what they literally said. Moreover, uncommon tasks with fewer datasets are more challenging, such as stance detection in the economic domain (most stance detection data is for political domain \citep{kuccuk2020stance}), moral exceptions, language inference under different cultures, and so on. With more fine-grained definitions on labels, LLMs have better performance in classification as seen from a higher F1 on GoEmotions than SemEvalT1 with more emotion classes defined. Additionally, more social context in the data can also result in better performance: for instance, they achieve a higher F1 on the CICERO dataset with both social situation description and dialogue data, than the SocialIQa dataset with only a simple description.

\paragraph{Performance Comparison with Humans.} Table \ref{tab:result} shows that, for every task there exists at least one LLM surpassing the average human performance. However, \textit{\textbf{LLMs perform worse than best human performance on most tasks on cognitive and situational intelligence}}. The gap between LLMs (e.g. GPT-4) and the best human performance is higher for more nuanced tasks (iSarcasm: 23.6 vs. SNIPS: 2.5), task in scarce domains (WTWT: 25.9 vs. SemEvalT6: 7.5) and more long-tailed situations (MoralExceptQA: 43.2 vs. NormBank: 10.9). On the other hand, \textit{\textbf{LLMs exceed average human performance on behavioral intelligence tasks}} with percentage preferred more than 50\% on all tasks. However, percentage preferred for LLMs (e.g. Claude) is lower in more dynamic and interactive situations (e.g. applying persuasion in dialogue: 74.3 vs. writing persuasive arguments: 93.6) with more constraints (e.g. with persona constraints: 64.0 vs. without persona constraints: 74.7) given. More qualitative analysis about human and LLMs performance is provided in Appendix \ref{sec:human_eval}.

\paragraph{Multiple intelligence.} LLMs in real-life social applications usually require multiple intelligence (e.g. interpreting intents under different cultural backgrounds) but they are still lacking in performance (CulturalNLI: 65.0). Table \ref{tab:result} shows they perform well for individual modules, so
systems can utilize LLMs for individual modules which LLMs do exceptionally well in and combine them organically to build a strong holistic system (e.g. combine emotion recognition and positive reframing components for a counseling system).

\section{Recommendations for the Future}
\label{sec:insights}
From analyzing current data landscape (\S \ref{sec:analysis}) and evaluating LLMs' performance (\S \ref{sec:model}), we unveil the most challenging aspects of social intelligence that remain unaddressed by existing data resources or model capabilities. Guided by insights from our results, we discuss possible future directions for dataset development below.

\subsection{Recommendations for Data Content}
Future datasets should focus more on \textit{\textbf{specific, nuanced, and long-tailed social situations}}. The LLMs we evaluated fell short on nuanced tasks like sarcasm and moral exceptions. As human expressions are diverse and subtle, and vary with complex linguistic contexts \citep{cruse2004meaning}, it is crucial to model the \textit{contextual complexity} to address the challenge of \textit{ambiguity} in language.
Long, multi-party interactions that go beyond conventional dialog or small groups are also critical for developing more sophisticated social intelligent systems that account for \textit{variation in discourse structure} and \textit{potential conflicts between individual and group objectives,} which produce novel social equilibria. 

As shown in Figure \ref{fig:in_type}, most datasets focus on just a single intelligence type. However, different intelligence types are not isolated but rather \textit{reciprocal} in real-world applications \citep{fan2022artificial}. Thus, there should be \textit{\textbf{more multifaceted datasets encompassing multiple intelligence types}} to promote holistic benchmarking. 

Moreover, there is a need for \textit{\textbf{better coverage on language, culture, countries, user groups,  and domain}}, as suggested by Figure \ref{fig:Lan} and LLMs' worse performance on stance detection in economic domain and culturally aware language inference. This will help models better generalize across populations and social contexts.
We need large and diverse datasets that can reflect the linguistic and cultural diversity of different user groups, which ensures that the models recognize and respond to social cue variations specific to different communities. Consider standard languages or mainstream user groups, versus low-resource languages and dialects, and vulnerable populations such as older adults or people with cognitive impairments.  

\subsection{Recommendations for Data Structure}
\label{subsec:social_ai_data_like}
There is a need for \textit{\textbf{higher interactivity in both data and evaluation}}. The field has an abundance of static resources and fewer interactive datasets. Our results show that dialogue context improves performance, and others have also argued that without interactivity, language models may be unable to fully develop key pillars of social intelligence, including Theory of Mind \citet{sap2022neural,bender2020climbing}. In this interactive settings, there will be an opportunity for a reduced focus on performance metrics like accuracy, and an increased focus on explainability, with socially intelligent AI systems that understand and can explain the factors that underlie their behaviors.

There can be significant shifts in people's values, beliefs and perspectives over time with societal changes like social movements, generational shifts and globalization. As a result, what social intelligence entails is constantly evolving. Thus, \textit{\textbf{data should undergo dynamic evolution}} to accurately capture social intelligence over different timeframes. Researchers can also consider a dynamic and flexible framework to allow future customization and extension \citep{zhou2023sotopia}.

Additionally, humans communicate using various modalities beyond language, such as gestures and facial expressions. Future datasets are encouraged to \textit{\textbf{incorporate multiple modalities}} to help AI systems develop a more accurate and well-rounded understanding of social contexts and social cues, leading to increased social intelligence.

\subsection{Recommendations for Data Collection}
Historically, social AI datasets have drawn heavily on randomly sourced crowdworkers who annotate datasets that have been scraped from social media or other online sources. There are at least three reasons why this paradigm will need to be replaced. 

The first concern is the issue of representation. A random sample of crowdworkers may not contain a fair representation of diverse viewpoints from a wide variety of sociodemographic backgrounds. Similar biases appear in randomly sampled social media data. Representation should extend beyond nationality to include diverse local regions, vulnerable populations, and people of different ages and genders. Both annotation and evaluation criteria should be designed in a way that \textit{\textbf{accounts for sociolinguistic variation}} and \textit{\textbf{considers diverging perspectives}}. Relatedly, crowdworkers may not be equipped to consistently identify subtle social cues. For this reason, we support \textbf{\textit{increasingly interdisciplinary, expert annotation efforts}} in which domain experts such as linguists, psychologists, anthropologists, and sociologists work to annotate high-quality social AI data resources.

Second, by passively observing decontextualized data, annotators may be unable to fully understand the social context behind any observed behaviors. There may be frequent misalignment between the behaviors expressed in random internet data and the lived experiences of the annotators. This motivates a more \textbf{\textit{active paradigm of dataset construction}} in which annotators \textit{participate} in the social interaction, and are thus \textit{de facto} experts on its situational context, any operational norms and cultural expectations that govern their behavior, as well as their own cognitive factors like personally motivating beliefs, intents, and emotions. As an added benefit, such active construction will naturally produce data with a high degree of interactivity (see \S\ref{subsec:social_ai_data_like}). 

Third, we encourage the field to closely consider how to effectively leverage LLMs to create \textit{\textbf{human-in-the-loop collaborative datasets}}, which applies both to the active generation of data previously mentioned and co-annotation of other social constructs \citep{li2023coannotating}.
Note that this differs from 
using LLMs to simulate synthetic social interaction data. In fact, we argue that it is still unclear whether simulation can produce high quality data with practical validity, since recent studies have shown caricatures and stereotypes in LLM-based simulations \cite{durmus2023towards,cheng2023compost}.  

Last but not least, we call for the development of annotation tools to facilitate the collection, visualization and annotation of different constructs in social intelligence, to allow for easy plug-in to existing crowdsourcing platforms and to support reproducible data collection. 

\subsection{Recommendations for Data Ethics}
Social AI datasets must be designed with ethical considerations, such as fairness, transparency, and privacy, to avoid perpetuating stereotypes or biases, and to respect user privacy. We envision that social AI dataset construction takes a community-centric approach where domain users co-design the tasks and data collection efforts with researchers (i.e., tasks of the community, by the community, and for the community), in addition to interdisciplinary collaboration among research fields.  This process will also benefit from protocols, compliance guides, culture- or country specific data use agreements to address any legal and ethical issues for creating and maintaining social AI datasets. 

\section{Conclusions}

We introduce \textit{Social AI Data infrastructure} with a theoretically grounded taxonomy and a data library of 480 NLP datasets, which facilitates standardization of the social intelligence concept in AI systems and organization of previous NLP datasets. We also conduct comprehensive analysis on the data library and evaluate LLMs' performance, offering insights on the current data landscape and future dataset development to advance social intelligence in NLP systems. It enables curation of high-quality datasets and holistic development of social intelligence in the NLP field.

\section*{Limitations}
\label{sec:limitations}
Although we try to be comprehensive, the datasets in our data library are not exhaustive as it is practically impossible to capture all datasets on socially intelligence. Moreover, we only crawled datasets on ACL Anthology, which is not representative of the whole academic space. As such, our analysis is more on relative comparison in the NLP domain rather than interpretation of the absolute figure. We encourage future work to further extend and contribute to our initial data library. Moreover, since LLMs have been trained on a large number of data, there may be data leakage issue where LLMs have seen some datasets in our experiment, making the performance reported higher than their actual capability. On top of that, since our work focuses more on obtaining insights about future dataset design (data aspect) instead of testing LLMs' social capability comprehensively (model aspect), we only select one simple and one nuanced dataset for each category for comparison purposes. Future work could leverage upon our infrastructure to design a comprehensive evaluation set for social intelligence to get insights on how models perform along each dimension of social intelligence.

\section*{Ethical Statement}
This study has been approved by the Institutional Review Board (IRB) at the researchers' institution, and we obtained participant consent with a standard institutional consent form.
One ethical concern is that models will become more capable of undesirable outcomes like persuasive misinformation or psychological manipulation as they become more socially intelligent. There may also be concerns that skilled anthropomorphic models will come to replace humans. These can not only lead to loss of trust in users \citep{mori2012uncanny} but also harm users' well-being \citep{salles2020anthropomorphism}. Our work proposes a standard concept and analysis the landscape and these risks are beyond the scope, but we acknowledge their presence and encourage future social AI data and systems to have clearer guidelines on the capabilities and limitations of AI systems to prevent deceptive and manipulative behaviours when advancing social intelligence.

\section*{Acknowledgements}
We are thankful to the members from SALT Lab for
their helpful feedback. Minzhi Li is
supported by the A*STAR Computing and Information Science (ACIS) Scholarship. Caleb Ziems is supported by the NSF Graduate Research Fellowship under Grant No. DGE-2039655.  This work was partially sponsored by NSF IIS-2247357 and IIS-2308994.
\bibliography{anthology,custom}
\nocite{le2019revisiting, sobhani2017dataset, li2021p, bohra2018dataset, qian2019benchmark, tadesse2019detection, jamil2017monitoring, forbes2020social, ziems2023normbank, jiang2022crecil, sap2019socialiqa, ghosal2022cicero, budzianowski2018multiwoz, zhu2020crosswoz, wang2019persuasion, habernal2016argument, li2017dailydialog, zhang-etal-2018-personalizing, bara2021mindcraft, litman2016teams, coucke2018snips, barbieri2020tweeteval, conforti2020will, mohammad2018semeval, mohammad2016semeval, demszky2020goemotions, sap2019socialiqa, ghosal2022cicero, ziems2023normbank, jin2022make, li2017dailydialog, zhang2018personalizing, habernal2016argument, wang2019persuasion, bertagnolli2020counsel, huang2023culturally}
\appendix
\section{Other Analysis}
\label{sec:appanalysis}
\paragraph{Perspective-Taking}
Perspective-taking is the act of considering an alternative point of view for the same situation, which is one aspect of social intelligence \citep{kosmitzki1993implicit}. 
Some work starts to pay more attention to two different perspectives which are \textit{\textbf{intended and perceived point of views}}. For example, \citet{oprea2019isarcasm} points out the difference between intended and perceived sarcasm from the perspectives of the author and audience, which is often overlooked in previous work. 
Thus, they asked for self-reported annotations 
to capture the intended sarcasm. The same also holds for other factors like emotion, which can be intended emotion by the author \citep{kleinberg2020measuring} or aroused emotion among the audience \citep{gambino2018distribution}.
\paragraph{Distribution of Data Sources} 
\begin{figure}[h!]
    \centering \includegraphics[width=\linewidth]{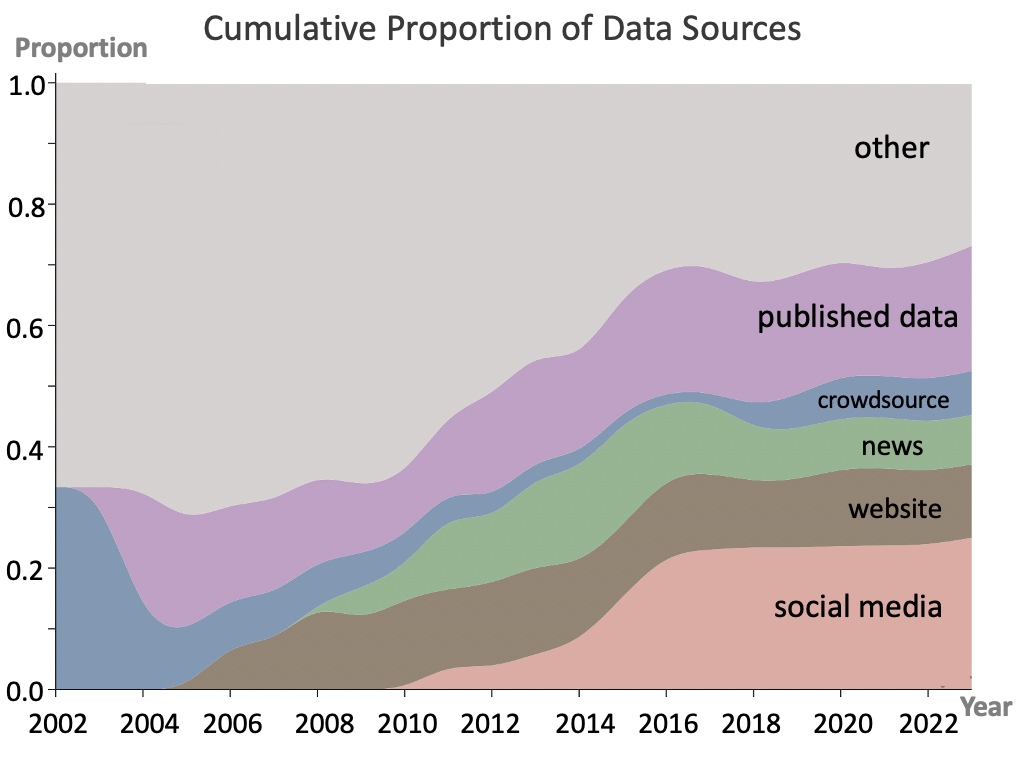}
    \caption{Percentage for major sources (social media, website, news, crowdsourcing, existing datasets) and other sources (e.g. book, speech etc.) over time.} 
    \label{fig:source}
\end{figure}
Most datasets in our data library use \textit{\textbf{social media}} as sources of data (see Figure \ref{fig:source}). The prevalence of data collection from social media has experienced \textit{\textbf{a significant surge from 2010}}. This might be due to an increase in the use of Twitter data \citep{baeth2019provenance}. 
In the meantime, relative proportions of traditional media like news and websites has experienced a decrease since then.

The second popular data source is \textit{\textbf{previously built data resources}}. New datasets leverage and extend upon previous ones in cases like translating to low-resource language \citep{ramaneswaran2022tamilatis}, introducing new evaluation criteria \citep{peng2020raddle} and adding new layers of annotation \citep{tigunova2021pride}.

\paragraph{Language and Culture Representation}
\begin{figure}[h!]
    \centering \includegraphics[width=\linewidth]{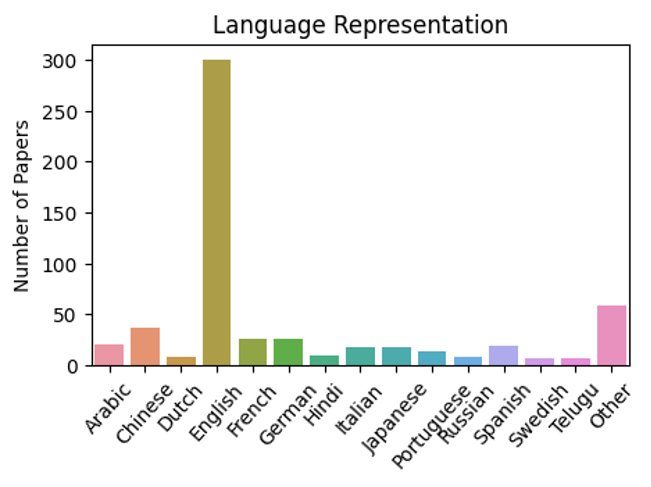}
    \caption{Distribution of datasets in different languages.} 
    \label{fig:Lan}
\end{figure}
Datasets surveyed in our data library covered up to 49 different types of languages. Figure \ref{fig:Lan} shows that \textit{\textbf{majority of study (62.5\%) uses English data}} to explore social intelligence and the number of such work is much higher than those in other languages. Moreover, there are more recent research efforts on code-mixing datasets about social intelligence, suggesting an increased representation of multilingual community \citep{chakravarthi2020sentiment, chakravarthi2020corpus, shetty2023poorvi}.

Additionally, most datasets in the data library (97.9\%) has unspecified cultural representation. However, the same sentence could have different meanings under different cultural contexts as social interpretations and social interactions vary from culture to culture. Therefore, there is a strong need for more future datasets with generations and annotations from different cultural backgrounds.

\begin{figure}[h!]
    \centering \includegraphics[width=\linewidth]{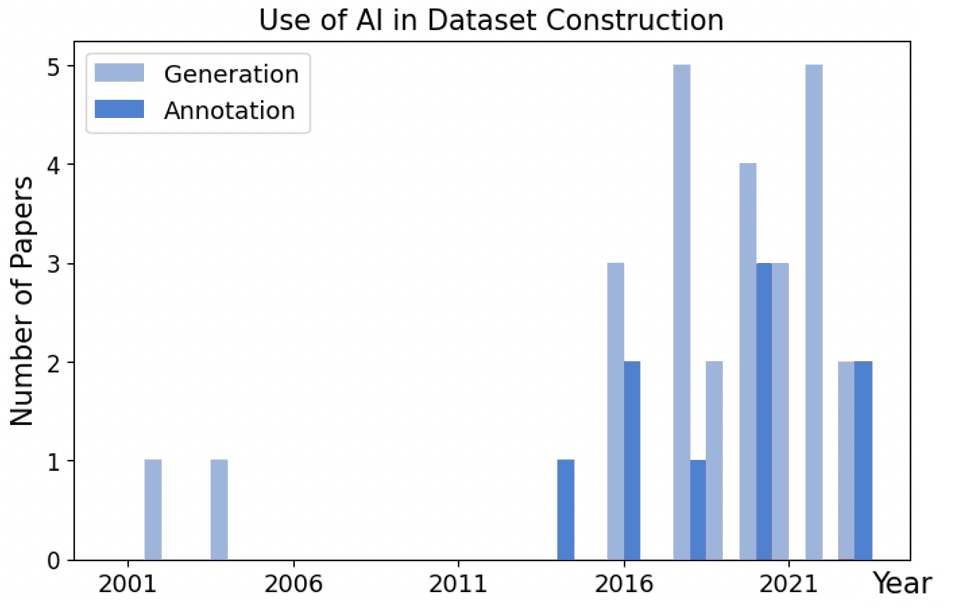}
    \caption{Number of papers that used AI for dataset generation or annotation.}
    \label{fig:useai}
\end{figure}

Overall, most of datasets contain textual content that is purely human-generated (95.0\%) and manually labeled (98.1\%). From Figure \ref{fig:useai}, we can see there is an \textit{\textbf{increasing trend in adoption of AI}} for generating and annotating datasets related to social intelligence. We can also see that number of work \textit{\textbf{using AI for generation is more than those for annotation}}. 

\paragraph{Incorporation of Different Modalities} 
\begin{figure}[h!]
    \centering \includegraphics[width=\linewidth]{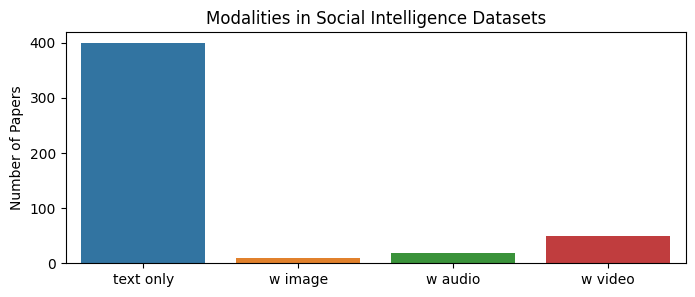}
    \caption{Distribution of modalities in datasets on social intelligence being surveyed.}
    \label{fig:modality}
\end{figure}
Because we only use crawled data from ACL Anthology, the majority datasets on social intelligence we surveyed are only in textual format. However, different modalities like image, audio and video can enhance learning of social intelligence with enriched social information embedded in other modalities.
\paragraph{More Open-sourced Community} 
\begin{figure}[h!]
    \centering \includegraphics[width=\linewidth]{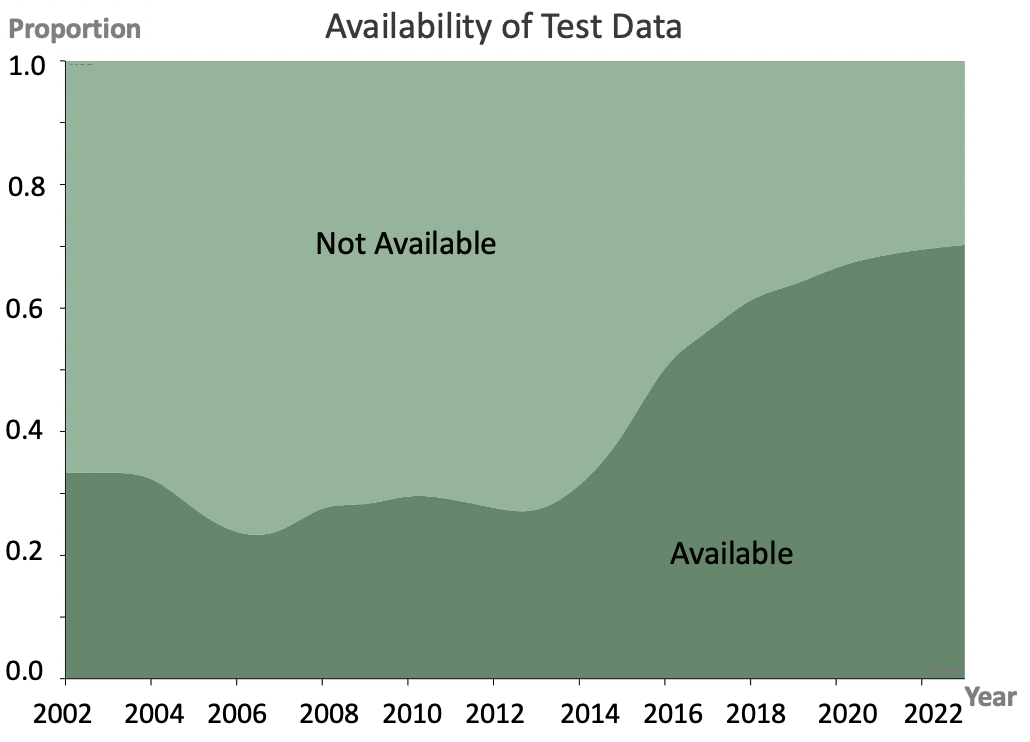}
    \caption{Proportion of available and unavailable social intelligence dataset (test set) over the years.}
    \label{fig:availability}
\end{figure}
Figure \ref{fig:availability} shows a promising trend where proportion of social intelligence datasets that are made available increases over the years. From 2014 onwards, the proportion of available has experienced a surge and there are \textit{\textbf{more publicly available social intelligence datasets}} than unavailable ones from 2016 onwards. Unavailable social intelligence datasets usually contain sensitive information such as mental illness \citep{santos2020searching, tasnim2023depac} and sexual orientation \citep{vasquez2023homo}. Additional measures like anonymizing, dara encryption and access control should be in place to protect data confidentiality while ensuring future work could have secure ways to use these data to advance research in sensitive domains.
\section{LLM Inference}

\paragraph{Model} We run inferences using Claude-v2, GPT-4-1106 and Llama2-13b models.

\paragraph{Prompt} We follow the recommended practice of prompting as described by \citet{ziems2023can}. We design prompts in MCQ format, give instructions after the context and clarify specific social concepts when necessary. 

\paragraph{Temperature Setting} We set temperature to 0 for classification tasks to ensure consistency and 0.7 for generation tasks to allow some diversity.
\section{Details of Human Evaluation}
\label{sec:human_eval}
\begin{table}[h!]
\centering
\def\arraystretch{1.8}
\resizebox{7.5cm}{!}{
\fontsize{18.88pt}{18.88pt}\selectfont
\begin{tabular}{@{}cc@{}}
\toprule
\textbf{Dataset}     & \textbf{Criteria}                                                                                                                             \\ \midrule
DailyDialogue        & \begin{tabular}[c]{@{}c@{}}information content, appropriateness, \\ engagement, naturalness, human-likeness\end{tabular}                      \\ \midrule
PersonaChat          & \begin{tabular}[c]{@{}c@{}}information content, appropriateness, engagement, \\ naturalness, human-likeness, persona consistency\end{tabular} \\ \midrule
Positive Reframing   & \begin{tabular}[c]{@{}c@{}}meaning preservation, degree of positivity,\\ naturalness, human-likeness\end{tabular}                             \\ \midrule
Counsel Chat         & \begin{tabular}[c]{@{}c@{}}information content, appropriateness, engagement, \\ naturalness, human-likeness, empathy\end{tabular}             \\ \midrule
Convincing Arguments & \begin{tabular}[c]{@{}c@{}}persuasiveness, information content\\ naturalness, human-likeness\end{tabular}                                     \\ \midrule
PersuasionforGood    & \begin{tabular}[c]{@{}c@{}}information content, appropriateness, engagement, \\ naturalness, human-likeness, persuasiveness\end{tabular}      \\ \bottomrule
\end{tabular}}
\caption{Criteria for different datasets used in human evaluation. They include general criteria like appropriateness and information content \citep{howcroft2020twenty} and task-specific criteria such as meaning preservation in positive reframing and persona consistency in persona controlled dialogue.}
\label{tab:criteria}
\end{table}
For interpretability, we also provide different criteria (see Table \ref{tab:criteria}) for each task and collect human's free-text explanation on top of an overall judgment. Below is the qualitative analysis for their free-text inputs for each dataset:
\paragraph{Daily Dialogue} A better responses is more specific and thoughtful and goes beyond the straightforward question a bit to make a personal anecdote. Some machine generated responses are too artificially friendly and sympathetic.

\paragraph{PersonaChat}
A response is better if it expresses more information (information content), responds directly to the question (appropriateness, naturalness) and expands on simple facts with personal details (naturalness, engagement). 

\paragraph{Positive Reframing} 
A better response is more \textit{multifaceted} and \textit{nuanced} by acknowledging all aspects of the input text. It better acknowledges and preserves the original meaning. Some machine generated responses feel forced and rated as worse.

\paragraph{Counsel Chat}
Better responses are more detailed, address the complexity of the situation and provide more specific and actionable advice.

\paragraph{ArgumentsPairs} 
Better arguments assert claims with evidence, provides specificity and contextualization of evidence and offer multiple complementary ideas that address different sub-concerns.

\paragraph{PersuasionforGood} 
Better responses are more specific and have a clear call to action. They better express emotional (pathos) and logical (logos) appeals. They take a \textit{genuine interest} in the listener.

In addition, we also collect crowdworkers' perceptions and comments on whether they think the text is generated by machine or human. 

We find that majority of people cannot correctly identify which response is generated by machine and a significant proportion choose the option \textit{`both generations are produced by human'}. Thus, more regulation and transparency on who generates the texts are needed since it is hard for people to distinguish human and machine generated texts. 

In general, they perceive more specific and thoughtful answers to be human generated and those brief and incomplete answers to be machine generated. Occasionally, some people hold an opposite view that more detailed and structured responses are from machine. This suggests for generation tasks, LLM has exceeded average human performance (see Table \ref{tab:result}) and it has also been acknowledged in \textit{human's perception} for a certain proportion of population.


\section{Keywords}
\label{sec:keyword}

\begin{table}[h]
\resizebox{\columnwidth}{!}{
\begin{tabular}{@{}ccc@{}}
\toprule
                                                                                  & \textbf{Title}                                                                                                                                                                                                                                                                                            & \textbf{Abstract}                                                                                                                                                                                                                                                                                                                                                                                                                                                                                                                                          \\ \midrule
\textbf{\begin{tabular}[c]{@{}c@{}}Dataset\\ Keywords\end{tabular}}               & \begin{tabular}[c]{@{}c@{}}'benchmark', 'corpus',\\ 'dataset' , 'annotat'     \end{tabular}                                                                                                                                                                                                                                                         & \begin{tabular}[c]{@{}c@{}}'introduce/build a dataset/benchmark', \\ 'introduce/build a largescale dataset/benchmark', \\ 'introduce/build a large-scale dataset/benchmark', \\ 'the first benchmark', \\ 'the first largescale benchmark', \\ 'the first large-scale benchmark'\end{tabular}                                                                                                                                                                                                                                                              \\ \midrule
\textbf{\begin{tabular}[c]{@{}c@{}}Social\\ Intelligence\\ Keywords\end{tabular}} & \multicolumn{2}{c}{\begin{tabular}[c]{@{}c@{}}‘cognitive’,‘sentiment’,‘agreement’,‘debate’,‘bargain’,\\‘commonsense’,‘emotion’, ‘polar’, ‘expressive’, \\‘argument’,  ‘autistic’, ‘information-providing’, ‘interaction’,\\ ‘opinion’,‘negotiation’, ‘dialog’, ‘affect’,\\ ‘public speaking’, ‘semeval’,‘stereotype’, ‘hate speech’, \\‘stance’, ‘persona’, ‘conversation’,‘communicative’,‘communicate’, \\ ‘recommendation’, ‘intent’, ‘communication’, ‘gender’, ‘age’,\\ ‘emotional’, ‘sympathy’,‘empathy’, ‘mental’, ‘norm’,\\ ‘culture/cultural’, ‘social relation’, ‘speaker’, ‘author profiling’,\\ ‘moral’, ‘ethic’, ‘privacy/secret’, ‘socially aware’, ‘prosocial’,\\ ‘moral’, ‘social situation’, ‘social context’, ‘social commonsense’, \\ ‘style’, ‘depression’, ‘anxiety’, ‘persuasion’, \\‘recommendation’, ‘theory of mind’, ‘audience’, ‘chat’\end{tabular}} \\ \bottomrule
\end{tabular}}
\caption{Keywords to filter for papers on dataset collection and social intelligence.}
\end{table}
\end{document}